\documentclass{iopjournal}

\usepackage[T1]{fontenc}
\usepackage{enumerate}
\usepackage{amssymb}
\usepackage{amsbsy}
\usepackage{bm}
\usepackage{mathtools}
\usepackage{physics}
\usepackage{siunitx}
\usepackage{bbm}
\usepackage{abraces}
\usepackage{mathrsfs}
\usepackage{xcolor}
\usepackage{mathrsfs}
\usepackage{bm}
\usepackage{hyperref}
\usepackage[capitalise]{cleveref}
\crefname{appendix}{App.}{Apps.}
\Crefname{appendix}{Appendix}{Appendices}

\makeatletter
\AddToHook{cmd/appendix/before}{\def\cref@section@alias{appendix}}
\AddToHook{cmd/appendix/before}{\def\cref@chapter@alias{appendix}}
\makeatother

\usepackage{booktabs}
\usepackage[
    backend=biber,
    style=phys,
    sorting=none,
    maxcitenames=5,
    maxbibnames=5
]{biblatex}
\newcommand{\im}{\mathrm{i}}
\newcommand{\up}{\uparrow}
\newcommand{\down}{\downarrow}
\newcommand{\expec}[1]{\langle #1 \rangle}

\newcommand{\vecstyle}[1]{\bm{#1}}

\newcommand{\vk}{\vecstyle{k}}
\newcommand{\vq}{\vecstyle{q}}

\newcommand{\opprod}[2]{\left( #1 \middle| #2 \right)}

\DeclareMathOperator{\hc}{H.c.}

\ExplSyntaxOn
\newcommand{\processmomenta}[1]{%
  \tl_set:Nn \l_tmpa_tl {#1}%
  \tl_replace_all:Nnn \l_tmpa_tl {k}{\vk}%
  \tl_replace_all:Nnn \l_tmpa_tl {q}{\vq}%
  \tl_use:N \l_tmpa_tl%
}
\ExplSyntaxOff

\NewDocumentCommand\ladder{O{}mm}{%
  \hat{c}_{%
    \processmomenta{#2},%
    #3%
  }^{%
    \IfNoValueTF{{#1}}{{\phantom{\dagger}}}{#1}%
  }%
}

\newcommand{\opSC}[1]{\ladder{-#1}{\down} \ladder{#1}{\up}}
\newcommand{\opSCd}[1]{\ladder[\dagger]{#1}{\up} \ladder[\dagger]{-#1}{\down}}

\newcommand{\opf}[2][{\phantom{\dagger}}]{\hat{f}_{#2}^{#1}}
\newcommand{\opn}[2]{\hat{n}_{#1,#2}}

\newcommand{\Ef}{E_\mathrm{F}}
\newcommand{\rhof}{\rho_\mathrm{F}}
\newcommand{\debye}{\omega_\mathrm{D}}

\newcommand{\dispersion}[1]{\varepsilon_{#1}}

\newcommand{\maxGap}{\Delta_\mathrm{max}}
\newcommand{\trueGap}{\Delta_\mathrm{true}}

\newcommand{\greenMatSymbol}{\mathcal{G}}
\newcommand{\greenMat}[1]{\greenMatSymbol_\mathrm{#1} (\omega)}
\newcommand{\spectralMatSymbol}{\mathcal{A}}
\newcommand{\spectralMat}[1]{\spectralMatSymbol_\mathrm{#1}  (\omega)}

\newcommand{\dynamicalMatrix}{\mathcal{M}}
\newcommand{\normMatrix}{\mathcal{N}}

\newcommand{\md}{\mathrm{d}}
\newcommand{\hamiltonian}{\mathcal{H}}

\newcommand{\opAmplitude}[1]{\hat{R}(#1)}
\newcommand{\opNumber}[1]{\hat{N}(#1)}
\newcommand{\opPhase}[1]{\hat{I}(#1)}

\begin{document}

\articletype{Paper} 

\title{Secondary Collective Excitations in Intermediate to Strong-Coupling Superconductors}

\author{Joshua Althüser$^1$\orcid{0009-0000-2109-2619} and Götz S. Uhrig$^{1}$\orcid{0000-0003-1961-0346}}

\affil{$^1$TU Dortmund University, Otto-Hahn Straße 4, 44227 Dortmund, Germany}


\email{joshua.althueser@tu-dortmund.de}
\email{goetz.uhrig@tu-dortmund.de}

\keywords{Superconductivity, Collective Excitations}

\begin{abstract}
Considering systematically derived energy-transfer-dependent effective electron-electron interactions leads to the appearance of secondary phase and amplitude modes in isotropic superconductors in the intermediate-to-strong-coupling regime.
We study the implications of such interactions on Bravais lattices by computing the corresponding response functions using the iterated equations of motion (iEoM) approach.
In the weak-coupling regime, we find the conventional, primary amplitude and phase modes at $\omega=2\Delta$ and $\omega=0$, respectively.
For intermediate coupling, the amplitude mode detaches from the quasiparticle continuum towards lower energies.
Increasing the coupling further leads to additional, long-lived secondary collective excitations below the continuum.
This phenomenon is largely independent of the underlying lattice and the specific Fermi level.
The amplitude and phase modes couple if the system is not particle-hole symmetric.
Additionally, we extend the method to compute eigenoperators, i.e., linear combinations of operators 
that excite each secondary mode specifically.
We identify nodal structures in the coefficients for these eigenoperators 
reminiscent of wave functions in the Hydrogen problem.
\end{abstract}

\section{Introduction}
\label{sec:introduction}

Superconductors have been at the heart of condensed matter research for over a century.
They represent important real-world applications due to their perfect electrical conductivity.
Still, many unanswered questions and unexplored areas remain in the realm of this phenomenon.
In this article, we explore the collective excitations in superconductors in particular.
We do this at zero temperature for simplicity.

For describing superconductivity, one commonly employs the same approximation as Bardeen, Cooper, and Schrieffer (BCS)~\cite{bardeen1957}  that the attraction between electrons is a constant in a small energy window around the Fermi edge, i.e.,
\begin{equation}
    g_\mathrm{BCS}(\vk, \vk') = -G \Theta(\debye-|\varepsilon_{\vk}-\Ef|)\Theta(\debye-|\varepsilon_{\vk'}-\Ef|),
\end{equation}
where $G$ is a coupling strength, $\Ef$ the Fermi energy, $\varepsilon_{\vk}$ the single-electron dispersion, and $\debye$ the Debye energy, which is used as a cutoff energy.
In this case, two distinct collective excitations exist: the amplitude (Higgs) mode related to the fluctuations of the 
magnitude of the order parameter $\Delta$, and the phase (Anderson-Bogoliubov) mode related to the fluctuations of the phase of $\Delta$. In this article, we refer to these two fundamental modes as the primary modes.

If one neglects the charged nature of the electrons, i.e., considers a neutral superfluid, the phase mode is the 
Goldstone boson of the broken $\mathrm{U}(1)$ phase symmetry~\cite{goldstone1961}.
In more rigorous treatments, including the charge, the phase mode couples to the electromagnetic degrees of freedom via the Anderson-Higgs mechanism, which lifts the mode's energy to the plasma frequency~\cite{anderson1958, kulik1981, schon1976}.
In contrast, the Higgs mode does not directly couple to the electromagnetic field.
In the weak-coupling limit, its energy is given by $2 \Delta$, i.e., it lies on the lower edge of the quasiparticle continuum.
Consequently, it can decay into quasiparticle states and dissipates~\cite{althuser2024, althuser2025, cea2014, dzero2024, fischer2018, hirashima1987, krull2016, measson2014, muller2019, reinhoffer2022, schmid1975, schwarz2020, sulaiman2024, tsuji2015, varma2002, volkov1973, yuzbashyan2006}.
By treating the interaction term in more detail, the Higgs mode shifts slightly below the quasiparticle continuum.
This implies that it becomes infinitely long-lived, i.e., dissipationless~\cite{lorenzana2024,lorenzana2024a,tian2026}.
Considering more elaborate momentum-dependent effective interaction potentials also makes the Higgs 
mode dissipationless. Concomitantly, the mode does not remain just below  $2 \Delta$. 
Instead, it shifts well into the gapped region free from particle-hole excitations~\cite{barankov2007,althuser2025,althuser2025a}.

In this article, we will work with an interaction derived from an electron-phonon interaction by a
continuous unitary transformation (CUT)\cite{kehrein2006, krull2012, lenz1996, mielke1997, mielke1997a}. 
In the pairing channel, it reads~\cite{krull2012}
\begin{equation}
\label{eqn:intro_g}
    g(\vk, \vk') = - \frac{|M_{\vk-\vk'}|^2}{\omega_{\vk-\vk'}} \Theta(\omega_{\vk-\vk'} - |\dispersion{\vk} - \dispersion{\vk'}|).
\end{equation}
Here, $M_{\vk-\vk'}$ is the electron-phonon coupling, and $\omega_{\vk-\vk'}$ is the phonon dispersion.
The advantage of this form of interaction is that retardation effects are inherently included as only electrons whose energy difference is smaller than $\omega_{\vk-\vk'}$ interact.
This means that the interaction occurs on a time scale $t > 1/\omega_{\vk-\vk'}$, which is the definition of retardation~\cite{kehrein2006}.

The interaction is derived perturbatively in $M$.
This means for \eqref{eqn:intro_g} to be justified $|M|\ll \debye$ should hold where $\debye$ is the phonon energy scale. 
Still, the dimensionless coupling $M^2 \rhof / \debye$ can be order unity for sufficiently large $\rhof$~\cite{yuzbashyan2022}, which is the electronic density-of-states (DOS) in the vicinity of the Fermi level.

In Ref.~\cite{althuser2025}, it was found that the nontrivial energy-transfer dependence interaction can induce
additional collective excitations with higher energy than the primary collective excitations described above.
These secondary modes appear in both the amplitude and the phase channel.
The aforementioned work investigated a rotationally invariant continuum system with parabolic kinetic energy
$\dispersion{\vk} \propto \vk^2$. This highly symmetric system allowed for the inclusion of long-range Coulomb interactions.
It was found that the Coulomb interaction merely affects the primary phase mode via the Anderson-Higgs mechanism~\cite{althuser2025, anderson1958, kulik1981, schon1976}.
All other modes are only affected indirectly by the diminishing of the superconducting gap function due to the Coulomb repulsion.

Our present goal is to perform calculations similar to those of Ref.~\cite{althuser2025} on \textit{lattice}
systems to understand how the secondary excitations depend on the system's geometry.
In particular, we investigate simple cubic (sc), body-centered cubic (bcc), and face-centered cubic (fcc) lattices.
To do so, we rewrite all occurring quantities in terms of the DOSes given explicitly in \cref{app:dos}.
The price to pay is that we refrain from including the Coulomb interaction.
Yet, its effects on the secondary modes are characterized mainly by the pseudo potential $\mu^*$ of Morel and Anderson 
$g \to g - \mu^*$, which effectively reduces the interaction strength~\cite{althuser2025, fischer2018, kostrzewa2018, mielke1997, sigrist2005, simonato2023}. Hence, all results obtained are relevant and justified, except for the precise value
of the interaction necessary for them to occur.

The remainder of this article is structured as follows.
In \cref{sec:model}, we introduce the model employed and discuss its static mean-field properties.
In \cref{sec:modes}, we go beyond the mean-field approximation and discuss the Fourier-transformed Green's functions in the amplitude and the phase channel.
We identify the collective excitations in the corresponding spectral functions.
Next, we introduce an algorithm that computes the specific linear combination of operators that excites a chosen collective excitation. 
We call these linear combinations eigenoperators of the mode because the resulting spectral functions 
display $\delta$ functions belonging only to the desired mode.
We refer to the coefficients defining the linear combinations of the eigenoperators as operator amplitudes.
In these amplitudes, we identify nodal structures reminiscent of wave functions for 
different principal quantum numbers in standard models such as the Hydrogen atom.
In \cref{sec:conclusion}, we conclude our results and provide possible directions for future studies.

\section{Model and static mean-field theory}
\label{sec:model}

We investigate the pairing Hamiltonian
\begin{equation}
    \label{eqn:h}
    \hamiltonian = \sum_{\vk \sigma} (\dispersion{\vk} - \mu) \ladder[\dagger]{\vk}{\sigma} \ladder{\vk}{\sigma} 
    + \frac{1}{N} \sum_{\vk \vk' \sigma} g(\vk, \vk') \ladder[\dagger]{\vk}{\sigma} \ladder[\dagger]{-\vk}{-\sigma} \ladder{-\vk'}{-\sigma} \ladder{\vk'}{\sigma},
\end{equation}
where $\ladder[(\dagger)]{\vk}{\sigma}$ annihilates (creates) an electron with momentum $\vk$ and spin $\sigma$,
$\dispersion{\vk}$ is the single-particle dispersion, $g(\vk, \vk')$ the interaction potential~\eqref{eqn:intro_g} allowing for superconductivity, and $N$ the number of lattice sites.
The chemical potential $\mu$ is chosen such that the system has the filling of the would-be normal state system 
for a given Fermi level $\Ef$. All calculations are done at $T=0$.
For all considered lattices, we choose the nearest-neighbor hopping elements such that $\dispersion{\vk} \in [-W, W]$.
This convention requires a global energy shift for the fcc lattice, but allows all considered cases to be treated 
on equal footing.
Throughout this article, we express all energies in units of half the bandwidth $W$.

For simplicity, we assume that $M_{\vk-\vk'} = M = \text{const}$ and that there is only a single Einstein phonon branch $\omega_{\vk-\vk'} = \debye$, so that the final interaction can be expressed in terms of the single-particle energies~\cite{althuser2025,althuser2025a}
\begin{equation}
\label{eqn:g}
    g(\varepsilon, \varepsilon') = - \frac{g}{2 \tilde{\rho}} \Theta(\debye - |\varepsilon - \varepsilon'|),
\end{equation}
where $g$ is a dimensionless coupling constant and
\begin{equation}
    \tilde{\rho} \coloneqq \frac{1}{2 \debye} \int_{\Ef - \debye}^{\Ef + \debye} \! \md \varepsilon \rho(\varepsilon)
\end{equation}
is the average DOS in the vicinity of the Fermi edge; the averaging is necessary to deal with the integrable 
divergence at the Fermi level in the bcc case at half-filling.
In this case, all quantities of interest merely depend on the single-particle energy.
Thus, we work in $\varepsilon$-space discretized into $N=16000$ equidistant points.

To gain access to the order parameter and to static expectation values, we apply a mean-field decoupling of the interaction term, yielding the mean-field Hamiltonian
\begin{equation}
\label{eqn:H_mf}
    \hamiltonian_\mathrm{MF} = \sum_{\vk \sigma} (\dispersion{\vk} - \mu) \ladder[\dagger]{\vk}{\sigma} \ladder{\vk}{\sigma} + \sum_{\vk} \left( \Delta(\varepsilon_{\vk}) \opSCd{\vk} + \hc \right)
\end{equation}
and the order parameter
\begin{equation}
\label{eqn:gap}
    \Delta(\varepsilon) =  \frac{g}{\tilde{\rho}} \int\!\md \varepsilon' \Theta(\debye - |\varepsilon - \varepsilon'|) \rho(\varepsilon') \frac{\Delta(\varepsilon')}{2 E(\varepsilon')}.
\end{equation}
We choose $\Delta(\varepsilon) \in \mathbb{R}$ fixing the global U(1) phase  to a zero without loss of generality
since the free choice of this phase constitutes a symmetry.
The quasiparticle dispersion is given by
\begin{equation}
    \label{eqn:quasiparticle}
    E(\varepsilon) = \pm \sqrt{(\varepsilon - \mu)^2 + |\Delta(\varepsilon)|^2}.
\end{equation}
The static mean field is determined self-consistently as usual.

\begin{figure}[!t]
    \centering
    \includegraphics{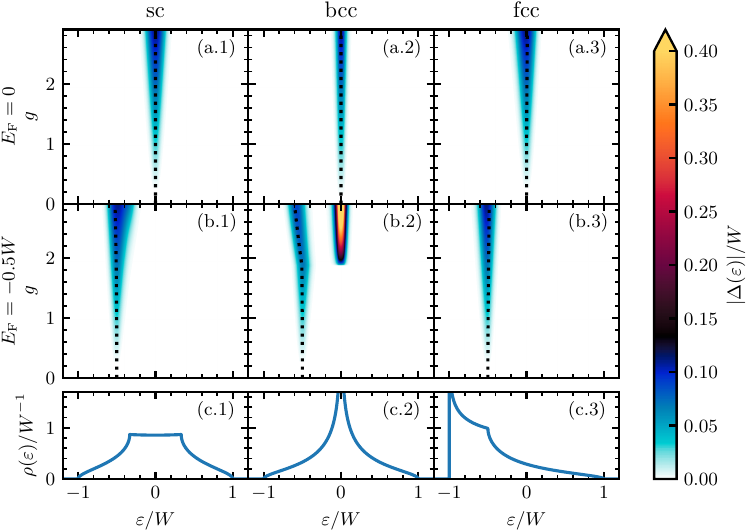}
    \caption{Plots of the gap function $\Delta(\varepsilon)$ (color scale) for different interaction strengths $g$ with the Fermi level set to (a) $\Ef = 0$ and (b) $\Ef = -0.5W$.
    The columns depict the results for the (1) sc, (2) bcc, and (3) fcc lattice.
    The black dotted lines mark the chemical potential $\mu$.
    The additional order parameter contribution in (b.2) has been discussed in detail in Ref.~\cite{althuser2025a}.
    The bottom row (c) shows the numerically evaluated DOS (cf. \cref{app:dos}).}
    \label{fig:gap_heatmap}
\end{figure}

The upper two rows (a, b) of \cref{fig:gap_heatmap} depict the self-consistently computed solution of the gap equation~\eqref{eqn:gap} for $\Ef = 0$ and $\Ef = -0.5W$, respectively.
The dimensionless coupling strength $g$ is varied along the $y$-axis.
The columns contain the data for the (1) sc, (2) bcc, and (3) fcc lattices.
The color scale denotes the magnitude of $\Delta(\varepsilon)$.
The black dotted lines mark the chemical potential $\mu$.
The bottom row (c) of \cref{fig:gap_heatmap} shows the corresponding DOS; its area is normalized to unity.
Note the integrable divergences at the lower boundary of the fcc lattice and at zero energy in the bcc lattice.
Generally, as $g$ is increased, $\maxGap \coloneqq \max_\varepsilon \Delta(\varepsilon)$ grows and 
$\Delta(\varepsilon)$ becomes finite for a larger interval of $\varepsilon$.

For $\Ef=0$, the sc (a.1) and bcc (a.2) systems display particle-hole symmetry, i.e., all quantities, such as $\rho(\varepsilon)$ and $\Delta(\varepsilon)$, are invariant under $\varepsilon \to -\varepsilon$.
In this case, $\mu~\equiv \Ef$. In all other cases, this statement does not hold, and the gap $\Delta(\varepsilon)$ is slanted towards larger DOS.
For instance, consider $\Ef = -0.5W$ on the sc lattice (b.1):
$\rho(\varepsilon < \Ef) < \rho(\varepsilon > \Ef)$ so that $\Delta(\varepsilon)$ is larger for $\varepsilon > \Ef$. 

The behavior of $\Delta(\varepsilon)$ is particularly peculiar for $\Ef = -0.5W$ on the bcc lattice.
For small $g$, the $\Delta(\varepsilon)$ is confined to the vicinity of the Fermi edge as expected.
At $g = 1.86$, however, $\Delta(\varepsilon)$ becomes finite also around $\varepsilon =0$ and $\Delta(0)$ quickly surpasses $\Delta(\Ef)$. 
This phenomenon was previously described and elucidated in Ref.~\cite{althuser2025a}. It results from the large accumulation of weight 
of the DOS in and around the logarithmic singularity.
This leads to an enhanced superconducting order, e.g., with significantly increased $T_c$.
Note, however, that the minimum of the quasiparticle dispersion is still located close to the Fermi edge.

\begin{figure}[!t]
    \centering
    \includegraphics{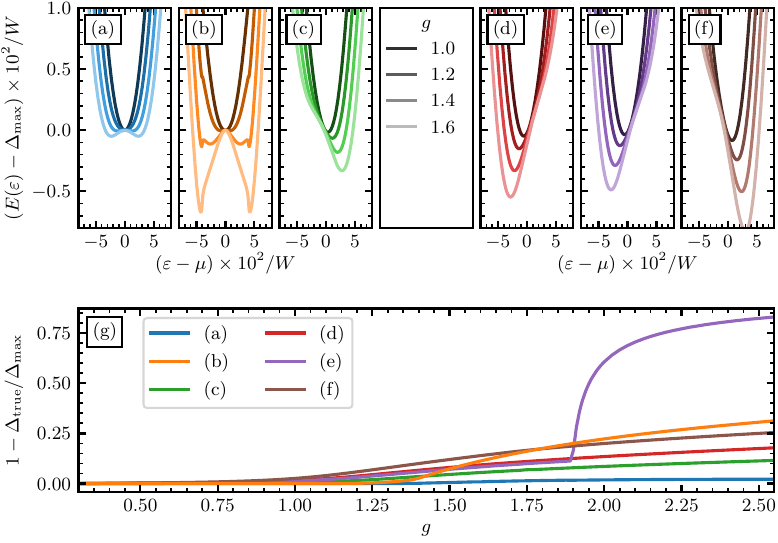}
    \caption{Plots of the quasiparticle dispersion~\eqref{eqn:quasiparticle} for the (a,d) sc, (b,e) bcc, and (c,f) fcc lattice.
    The left group (a--c) displays the data for $\Ef =0$, while the right group (d--f) displays the data for $\Ef=-0.5W$.
    Lines of different brightness represent different interaction strengths $g$ as indicated in the legend.
    Panel (g) depicts the system's true energy gap relative to the maximum value of the gap function.
    The six lines of different colors correspond to the cases shown in the upper panels.}
    \label{fig:max_true}
\end{figure}

Let us next discuss the shape of the quasiparticle dispersion~\eqref{eqn:quasiparticle}.
Typically, its minimum is located at $\varepsilon_\mathrm{min} = \mu$.
Yet, this need not be the case in general.
If the order parameter $\Delta(\varepsilon)$ falls off faster than $|\varepsilon - \mu|$ rises upon varying $\epsilon$ away from 
$\mu$, the minimum of $E(\varepsilon)$ is shifted~\cite{althuser2025}.
We investigate this behavior in more detail in \cref{fig:max_true}.
The upper row depicts the quasiparticle dispersion shifted by $\maxGap \coloneqq \max_\varepsilon \Delta(\varepsilon)$ 
in the close vicinity of $\varepsilon = \mu$ for various coupling strengths $g$.
We set $\Ef = 0$ for panels (a--c) and $\Ef = -0.5W$ for panels (d--f).
Again, the results for the sc (a, d), bcc (b, e), and fcc (c, f) lattices are displayed.
The bottom panel (g) shows the ratio of the true gap $\trueGap \coloneqq \min_\varepsilon E(\varepsilon)$ to the maximum value $\maxGap$.

For sufficiently small $g$, we find $\varepsilon_\mathrm{min} = \mu$, and $\trueGap = \maxGap$ in all cases.
Distinct deviations, however, occur for larger values of $g$.
The particle-hole symmetric sc lattice (a) presents the simplest case.
Here, two side minima emerge and become more pronounced as $g$ increases.
For the bcc lattice with $\Ef=0$ (b), $\Delta(\varepsilon)$ falls off rapidly for $\varepsilon \neq 0$ due to the strong variation of the DOS.
Therefore, the side minima emerge faster than in the sc case.
Additionally, there are two distinct side minima for a small parameter range, see the line for $g=1.4$.
The outer minimum becomes the global one at $g \approx 1.375$.

The remaining four cases (c--f) behave qualitatively the same.
There is no particle-hole symmetry in these systems.
For $\varepsilon \neq \mu$, the DOS rises in one direction and falls off in the other, which translates
to $\Delta(\varepsilon)$ as well.
For the fcc lattice (c, f), the order parameter reaches its maximum for $\varepsilon < \mu$, 
which causes the minimum of $E(\varepsilon)$ to shift to $\varepsilon_\mathrm{min} > \mu$. 
This behavior is reversed on the sc and bcc lattices for $\Ef=-0.5W$ (d, e).

As can be inferred from \cref{fig:max_true} (g), the difference between the true gap $\trueGap$ and the maximum of the order parameter $\maxGap$ rises similarly in almost all cases.
The only striking exception is the bcc lattice with $\Ef=-0.5W$ (e). This stems from the 
emergent peak in $\Delta(\varepsilon)$ around $\varepsilon=0$
which surmounts $\Delta(\varepsilon \approx \Ef)$~\cite{althuser2025a}, see panel (b.2) in
Fig.~\ref{fig:gap_heatmap}.

\section{Collective excitations}
\label{sec:modes}

The simple mean-field approximation cannot capture collective excitations because the mean-field Hamiltonian effectively describes a single-particle problem. Therefore, we turn to the iterated equations of motion (iEoM) approach.
This method reduces the Heisenberg equation of motion for the operators of a suitable operator basis $\mathfrak{B}$ to an algebraic system of differential equations by applying an operator (pseudo-)scalar product to both sides of the equation~\cite{althuser2024, althuser2025, althuser2025a, hamerla2013, hamerla2014, schwarz2020a, uhrig2009}. 
Conceptually, the idea has been employed in a wide array of studies; see, for example, Refs.~\cite{lelwalagamacharige2019, catalano2021, avella2012, mancini2004, roth1969, tserkovnikov1981}.

In particular, we closely follow the procedure presented in Ref.~\cite{althuser2024}.
To do so, we choose the operator pseudo-scalar product
\begin{equation}
    \opprod{\hat{A}}{\hat{B}} \coloneqq \expec{[\hat{A}^\dagger, \hat{B}]},
\end{equation}
where the expectation value is taken with respect to the mean-field Hamiltonian~\eqref{eqn:H_mf}, i.e., its ground state.
The initial task is to compute the dynamical matrix
\begin{subequations}
\begin{equation}
\label{eqn:dynmat}
    \dynamicalMatrix_{ij} = \opprod{\hat{O}_i}{[\hamiltonian, \hat{O}_j]}
\end{equation}
and the norm matrix
\begin{equation}
    \normMatrix_{ij} = \opprod{\hat{O}_i}{\hat{O}_j}
\end{equation}
\end{subequations}
for each $\hat{O}_i \in \mathfrak{B}$.

In our study, $\mathfrak{B}$ contains the operators
\begin{subequations}
\label{eqn:basis_operators}
\begin{align}
    \opAmplitude{\varepsilon} &= \frac{1}{\sqrt{N}} \sum_{\vk} \delta(\dispersion{\vk} - \varepsilon) \left( \opSCd{\vk} + \opSC{\vk} \right) \\
    \opNumber{\varepsilon} &= \frac{1}{\sqrt{N}} \sum_{\vk} \delta(\dispersion{\vk} - \varepsilon) \left( \ladder[\dagger]{\vk}{\up} \ladder{\vk}{\up} + \ladder[\dagger]{\vk}{\down} \ladder{\vk}{\down} \right) \\
    \opPhase{\varepsilon} &= \frac{1}{\sqrt{N}} \sum_{\vk} \delta(\dispersion{\vk} - \varepsilon) \left( \opSCd{\vk} - \opSC{\vk} \right)
\end{align}
\end{subequations}
for all values of $\varepsilon$.
We stress that the commutator in \cref{eqn:dynmat} must be computed with respect to the full Hamiltonian~\eqref{eqn:h} to capture the collective behavior. This is the crucial step that goes beyond the mean-field approximation. 
The expectation values of the resulting operator terms are again evaluated with respect to the mean-field Hamiltonian, so that Wick's theorem can be applied.

The algorithm presented in Ref.~\cite{althuser2024} links $\dynamicalMatrix$ and $\normMatrix$ to the system's Fourier transformed Green's functions
\begin{equation}
    \greenMatSymbol_{\alpha}(z=\omega + \im 0^+) = - \im \int_0^\infty \! e^{\im z t} \expec{[\mathfrak{A}_\alpha (t), \mathfrak{A}_\alpha^\dagger]} \md t.
\end{equation}
In particular, we are interested in the Green's functions of the operators~\cite{althuser2025}
\begin{subequations}
\begin{align}
    \mathfrak{A}_\mathrm{Higgs} &=   \frac{1}{\sqrt{N}} \sum_{\vk} \left( \opSCd{\vk} + \opSC{\vk} \right) = \int \! \opAmplitude{\varepsilon} \md \varepsilon   \\
    \mathfrak{A}_\mathrm{Phase} &= \frac{\im}{\sqrt{N}} \sum_{\vk} \left( \opSCd{\vk} - \opSC{\vk} \right) = \im \int \! \opPhase{\varepsilon} \md \varepsilon ,
\end{align}
\end{subequations}
which excite the amplitude and phase fluctuations, respectively.
We emphasize that these operators comply with the point symmetries of the underlying lattice. 
Hence, the observed collective excitations do not have a finite angular momentum associated with them.
Rather, they have (extended) $s$-wave character in the usual terminology.
As far as the translational symmetry is concerned, the operators act at zero momentum, i.e., they
do not add or subtract momentum. Thus, the corresponding collective excitations live at zero momentum.

\subsection{Spectral functions}

\begin{figure}[!t]
    \centering
    \includegraphics{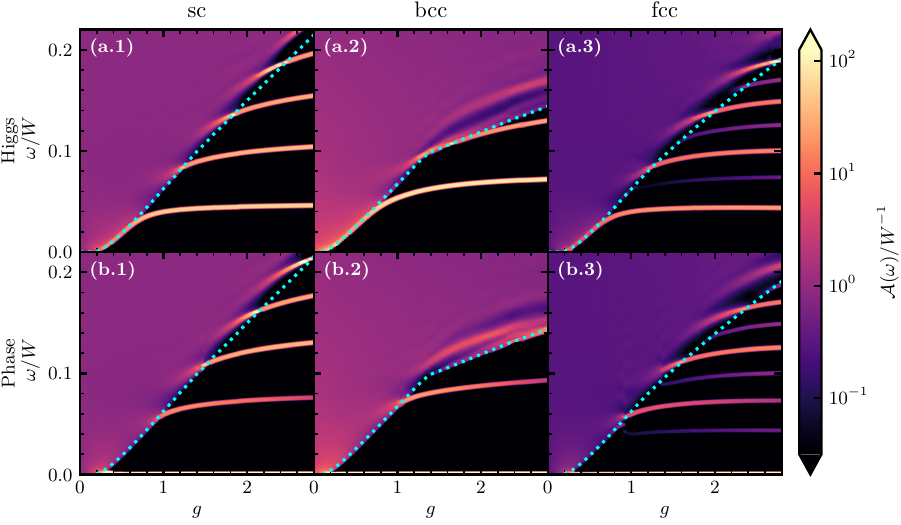}
    \caption{Plot of the computed spectral functions. 
    The top row (a) depicts the spectral function $\spectralMat{Higgs}$, while the bottom row depicts $\spectralMat{Phase}$.
    The individual columns contain the results for the three lattices, from left to right (1) sc, (2) bcc, and (3) fcc.
    The color scale represents the magnitude of the spectral functions.
    The $x$-axis indicates the interaction strength $g$.
    Mathematically, the subgap peaks are $\delta$-peaks, which we represent by Gaussian bell curves of the same weight and the 
    artificial width $\sigma=0.0005W$.
    The peak at $\omega=0$ in $\spectralMat{Phase}$ is an exception because
    it corresponds to the derivative of a $\delta$-peak and is therefore plotted by the derivative of a Gaussian bell.
    The cyan line indicates the lower edge of the quasiparticle continuum $2 \trueGap$.
    The color scale is logarithmic to highlight the individual modes better.
    The Fermi level is set to $\Ef=0$ in all panels.
    }
    \label{fig:heatmap_g}
\end{figure}

In this section, we study the spectral functions 
$\spectralMat{\alpha} = - (1/\pi) \Im [\greenMatSymbol_{\alpha}(\omega + \im 0^+)]$.
Since the exciting operators are of bosonic character, being built from pairs of fermionic
operators, the spectral functions are odd in $\omega$; hence, one can restrict the discussion to 
positive frequencies.
Let us begin the discussion of these functions for $\Ef = 0$.

The upper row (a) of \cref{fig:heatmap_g} depicts $\spectralMat{Higgs}$, while the lower row (b) of \cref{fig:heatmap_g} depicts $\spectralMat{Phase}$ for the (1) sc, (2) bcc, and (3) fcc lattices.
The interaction strength $g$ is varied along the $x$-axis.
The color scale indicates the magnitude of the spectral functions.
Subgap peaks with a finite energy are mathematically $\delta$-peaks and are represented by Gaussian bells with the same weight and width $\sigma = 0.0005W$.
The phase spectral functions additionally display a sharp antisymmetric structure at $\omega=0$. Mathematically it is 
the derivative of a $\delta$-function, which we depict by the derivative of a Gaussian bell curve.
This behavior is typical for Goldstone modes identified by our approach~\cite{althuser2024,althuser2025}.
This particular excitation is the Goldstone mode resulting from the broken $\mathrm{U}(1)$ phase symmetry of 
the system~\cite{goldstone1961}.
If the Coulomb interaction were included, this mode would be shifted to higher energies, in accordance with the Anderson-Higgs mechanism~\cite{anderson1958, schon1976, kulik1981,althuser2025}.

For weak coupling, we find the well-known Higgs mode in $\spectralMat{Higgs}$ at $\omega_\mathrm{Higgs} = 2 \maxGap = 2 \trueGap$~\cite{althuser2024, althuser2025, cea2014, dzero2024, fischer2018, hirashima1987, krull2016, measson2014, muller2019, reinhoffer2022, schmid1975, schwarz2020, sulaiman2024, tsuji2015, varma2002, volkov1973, yuzbashyan2006}. 
In all cases presented here, this excitation detaches itself from the quasiparticle continuum, i.e., $\omega_\mathrm{Higgs} < 2 \trueGap$, 
as the interaction strength is increased.
This behavior has been described in previous studies, where the Higgs mode was observed to become
infinitely long-lived, i.e., dissipationless, because it is located below the two-particle continuum
at zero temperature. Hence, there is no decay channel.
Here, the detaching is driven by the nontrivial momentum dependence of the interaction~\cite{barankov2007,althuser2025}.
In principle, the Higgs mode can also detach itself for large constant interactions.
This result can then be achieved by considering corrections arising from quantum fluctuations~\cite{tian2026}.
In this case, however, the mode remains close to  the top of the gapped region $\omega_\mathrm{Higgs} \lesssim 2 \Delta$~\cite{lorenzana2024, lorenzana2024a}, which evidently does not hold here.

Increasing $g$ further causes \textit{secondary} modes to emerge from the quasiparticle continuum.
On the sc (1) and bcc (2) lattices, they appear in an alternating fashion first in $\spectralMat{Phase}$ and then in $\spectralMat{Higgs}$.
These modes emerge at regular intervals, which we investigate in more detail below.
All in all, the present situation is qualitatively the same as in Ref.~\cite{althuser2025}, where a parabolic dispersion and the Coulomb repulsion were considered.

The fcc lattice (3) displays another feature due to the lack of particle-hole symmetry.
The phase and Higgs modes appear in \textit{both} spectral functions.
We will stick to the previous naming scheme: the lowest finite-energy excitation is the primary Higgs mode; the next is the secondary phase mode; then the secondary Higgs mode, and so on.
This identification is consistent with the fact that the Higgs modes have more weight in 
$\spectralMat{Higgs}$, while the phase modes have more weight in $\spectralMat{Phase}$.
For instance, the lowest peak in $\spectralMat{Higgs}$, i.e., the primary Higgs mode, lies at the same energy 
as the lowest mode in $\spectralMat{Phase}$. Thus, both represent the same collective excitation.
But the former peak is much more pronounced than the latter one, which is a consequence of different
matrix elements by which the collective modes are excited.
We attribute this behavior to the absence of a particle-hole symmetry in this system.
It is known that, in the standard BCS theory, the Higgs and the phase channels couple to each other for 
$\omega \neq 0$ if the system does not exhibit this symmetry~\cite{engelbrecht1997,kos2004,fischer2018}.
Thus, we confirm that this effect is not specific to the BCS approximation, but
applies to the interaction~\eqref{eqn:intro_g} as well.

\begin{figure}[!t]
    \centering
    \includegraphics{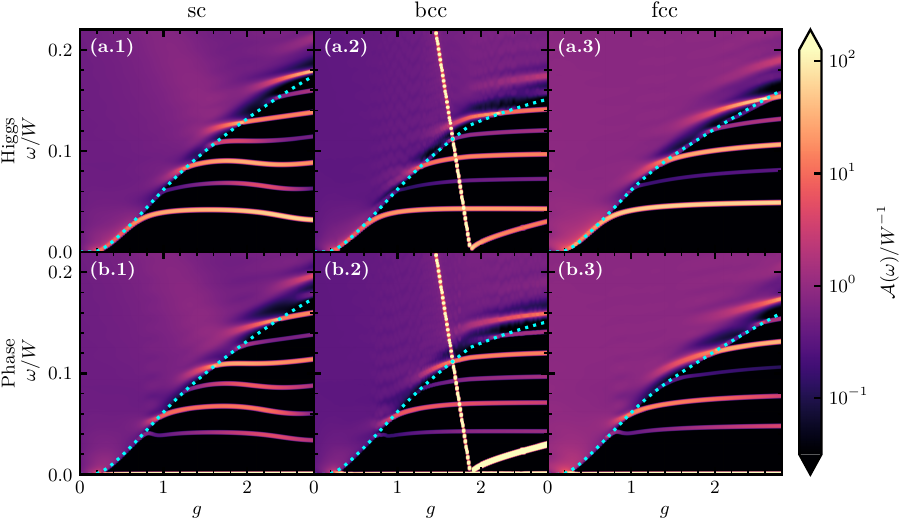}
    \caption{Same as \cref{fig:heatmap_g}, but with $\Ef = -0.5W$.}
    \label{fig:heatmap_g_EF}
\end{figure}

Next, we consider $\Ef = -0.5W$ in \cref{fig:heatmap_g_EF}.
The figure is structured in the same way as the previous one.
First and foremost, we again observe the detachment of the primary Higgs mode 
from the lower boundary of the continuum and the emergence of secondary modes.
Due to the finite Fermi level, the systems are not at half-filling, and none of them
exhibits particle-hole symmetry anymore.
Hence, we find signatures of the modes in both spectral functions as discussed above
for the case of the fcc lattice.
All in all, the present analysis indicates that the existence of these secondary modes is 
a generic feature of large coupling strengths, independent of the underlying lattice geometry or the Fermi level.

The emergence of the third amplitude mode on the fcc lattice is somewhat peculiar.
It emerges around $g \approx 2.2$, but sticks close to the continuum boundary 
until the fourth phase mode is about to emerge.
Then, both modes experience an anticrossing. We observed this special kind of behavior only here.
It is described in more detail in \cref{app:third_fcc_mode}.

Additionally, we find another unconventional excitation on the bcc lattice (2).
It is quite sharp even inside the quasiparticle continuum and becomes soft at $g\approx 1.86$.
This is precisely the interaction strength at which the additional order parameter contribution 
in \cref{fig:gap_heatmap}(b.2) appears. This phenomenon has been described and analyzed 
comprehensively in the Ref.~\cite{althuser2025a}.
The feature is linked to the large weight accumulation in the electronic DOS away from the Fermi level.
The mode is sharp even inside the continuum because it is related to quasiparticle states of the additional order.
These states do not lie close to the Fermi level, but at high energies. 
Within the used approximation, they are not directly coupled to the low-energy states at the Fermi level.
If the additional order is directly coupled to the conventional one, for instance, due to additional interactions, 
this mode is broadened, but still persists~\cite{althuser2025a}.

\begin{figure}[!t]
    \centering
    \includegraphics{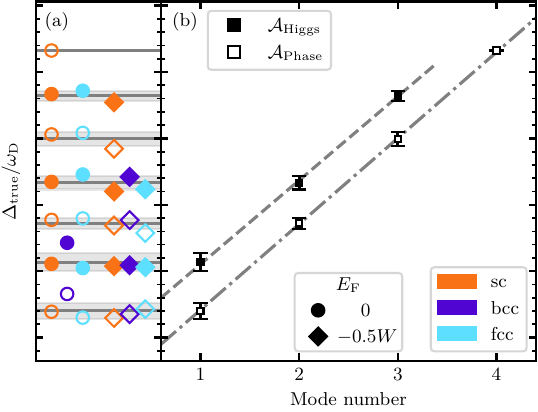}
    \caption{True gap $\trueGap$ in units of the Debye frequency $\debye$ at which the secondary modes emerge from the quasiparticle continuum.
    The filled markers correspond to Higgs modes, while the open ones correspond to phase modes.
    Panel (a) displays the data for the six investigated cases (three lattices, each at two Fermi levels).
    The colors correspond to the lattices, while the marker shape corresponds to different Fermi energies $\Ef$.
    The gray lines mark the averages, while the shading indicates the standard deviation.
    In panel (b), we show these averages as a function of the mode number.
    The gray lines are linear fits of the data.
    The distance in $\trueGap$ between the emergence of additional modes is given by the slope of the fit.
    The slopes are $s_\mathrm{Higgs} = 0.63 \pm 0.04$ and $s_\mathrm{Phase} = 0.65 \pm 0.05$.
    }
    \label{fig:mode_emergence}
\end{figure}

In \cref{fig:mode_emergence}, we analyze when the secondary modes emerge from the quasiparticle continuum.
The $y$-axis displays the energy gap $\trueGap$ in units of $\debye$ at which the modes first emerge, i.e.,
split off the lower edge of the continuum.
The markers in panel (a) represent the data of the individual cases studied in this article.
The colors correspond to the three lattices and the marker shape to the two considered Fermi levels.
Filled markers represent Higgs modes while open markers represent phase modes.
The gray lines depict the average $\trueGap / \debye$ for any given mode, and the gray shading marks the standard deviation of the average.
Remarkably, the modes emerge at about the same energy, independent of the details, except for minor deviations.
The comparatively largest deviations occur for the bcc lattice with $\Ef=0$, 
which we attribute to the logarithmic divergence of the DOS.

In panel (b), the averages are displayed as a function of the mode number and fitted linearly.
The thereby obtained slopes $s$ determine the approximate amount by which the energy gap must grow 
for an additional mode to appear.
The slopes are $s_\mathrm{Higgs} = 0.63 \pm 0.04$ and $s_\mathrm{Phase} = 0.65 \pm 0.05$.
As these values are equal within their uncertainties, we combine them and take their average as $s = 0.64 \pm 0.03$.
For the continuum systems investigated in Ref.~\cite{althuser2025}, one similarly obtains $s = 0.636 \pm 0.006$.
Importantly, the results in Ref.~\cite{althuser2025} indicate that the slopes appear to be largely independent 
of the Coulomb interaction.
The fact that this slope is essentially the same as the present result provides evidence for the universality of the emergence of the secondary Higgs and phase modes.

\subsection{Amplitudes of the specific mode-exciting operators}
\label{sec:amplitude_discussion}

Besides identifying the existence of collective excitations, it is highly informative which kinds of operators are necessary to excite them, especially in view of potential experimental verifications.
Therefore, we compute the linear combination of operators that excites only one specific mode
at a time. We call these linear combinations \textit{eigenoperators}.
The reason behind this name will become clear shortly.
We denote the eigenoperator of the $n$-th Higgs mode by
\begin{equation}
\label{eqn:excite_higgs}
    \mathfrak{A}_\mathrm{Higgs}^{(n)} = \sum_{j=1}^N {\alpha}_{j}^{(n)} \opAmplitude{\varepsilon_j} + 
    \sum_{j=1}^N {\nu}_j^{(n)} \opNumber{\varepsilon_j},
\end{equation}
where $\hat{R}{(\varepsilon_j)}$ and $\hat{N}{(\varepsilon_j)}$ are the operators from \cref{eqn:basis_operators}.
Analogously, the $n$-th phase mode is excited by
\begin{equation}
\label{eqn:excite_phase}
    \mathfrak{A}_\mathrm{Phase}^{(n)} = \sum_{j=1}^N \psi_j^{(n)} \opPhase{\varepsilon_j},
\end{equation}
where $\opPhase{\varepsilon_j}$ is defined in \cref{eqn:basis_operators}.
We call the coefficients $\alpha_j^{(n)}$, $\nu_j^{(n)}$, and $\psi_j^{(n)}$ \textit{operator amplitudes}.
The algorithm to compute these amplitudes is presented and derived in \cref{app:computing_operators}.

We point out that the overall normalization of  
\begin{equation}
\vecstyle{\chi}^{(n)} = \begin{pmatrix} \vecstyle{\alpha}^{(n)} \\ \vecstyle{\nu}^{(n)} \end{pmatrix}
\end{equation}
is arbitrary because it only enters as a global factor in the Green's functions.
For plotting purposes, we fix this factor such that $\max_j |{\chi}_j^{(n)}| = 1$.
Note, however, that the ratio $|\vecstyle{\alpha}^{(n)}| / |\vecstyle{\nu}^{(n)}|$ is not arbitrary 
and is computed by the algorithm described in \cref{app:computing_operators}.

The Fourier-transformed Green's function with respect to an eigenoperator simply reads
\begin{equation}
    \greenMat{mode} = \frac{|C|^2}{z^2 - \omega_\mathrm{0}^2},
    \label{eq:eigenop}
\end{equation}
where $C$ is the aforementioned global factor, 
$z\coloneqq \omega + \im 0^+$, and $\omega_\mathrm{0}$ is the mode's energy.
For interpretation, the analogy to a harmonic oscillator, i.e., a simple bosonic system 
$\hamiltonian = \omega_0 \hat{b}^\dagger \hat{b}$ is helpful.
Computing the Green's function with respect to $\mathfrak{A} \coloneqq (\hat{b}^\dagger + \hat{b})$ 
leads to the same structure \eqref{eq:eigenop}.
Such an operator has the property $[\hamiltonian, [\hamiltonian, \mathfrak{A}]] = \omega_0^2 \mathfrak{A}$.
Hence, the $\mathfrak{A}$ is an eigenoperator of the double-commutation with the Hamiltonian.
This structure gives rise to the two poles at $\pm \omega_0$ in the Green's function.
The eigenoperators we compute here for the superconducting system are essentially such operators $\mathfrak{A}$.
Although the superconducting system is much more complex, the analogy fully applies within the used approximations.
Thus, the eigenoperator $\mathfrak{A}$ for a given mode behaves like the sum of the bosonic creation 
and annihilation operators corresponding to the energy $\omega_\mathrm{0}$.

\begin{figure}[!t]
    \centering
    \includegraphics{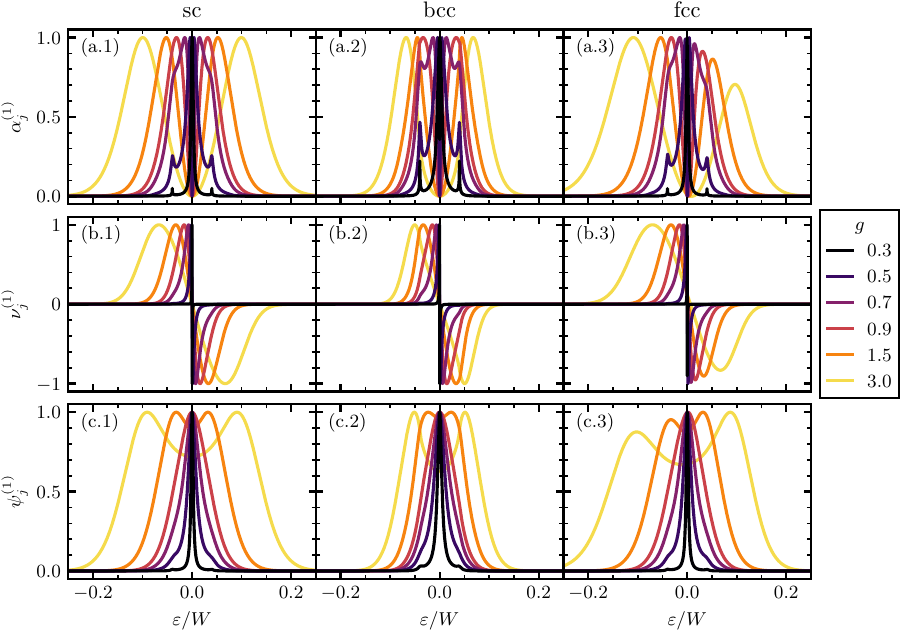}
    \caption{The operator amplitudes (see \cref{eqn:excite_higgs,eqn:excite_phase}) 
    for the primary Higgs and phase modes on the (1) sc, (2) bcc, and (3) fcc lattice with $\Ef = 0$.
    The different colors correspond to different $g$ values, see legend.
    The Higgs mode is excited by a combination of $\opAmplitude{\varepsilon}$ (a) and $\opNumber{\varepsilon}$ (b)
    while the phase mode only requires $\opPhase{\varepsilon}$ (c).
    Consequently, the coefficients in (a) and (b) are linked.
    Still, we normalize them such that the maximum in each panel is $1$ to improve visibility.
    The total contributions of the pair creation operators (see \cref{eqn:cpair}) and of the number operators (see \cref{eqn:cnum}) are shown in \cref{fig:primary_mode_integrals}.
    }
    \label{fig:primary_operators}
\end{figure}

For brevity, we only discuss the case $\Ef = 0$ here.
The cases $\Ef = -0.5W$ behave qualitatively the same as the fcc lattice 
for $\Ef = 0$ because the most important difference is the absence of particle-hole symmetry.
We first consider the primary modes.
Their corresponding operator amplitudes are plotted in \cref{fig:primary_operators}.
The columns depict the results for the three different lattices.
The rows depict the operator amplitudes of (a) $\opAmplitude{\varepsilon}$, (b) $\opNumber{\varepsilon}$, and (c) $\opPhase{\varepsilon}$. The colors correspond to different values of $g$.
For small values of $g$, one kind of amplitude dominates over the other
$\max_j \alpha_j^{(1)} \ll \max_j \nu_j^{(1)}$.
For good readability, we therefore normalize each curve in each panel 
to have a maximum value of $1$. We will discuss the physical meaning of the
strongly differing amplitudes for small $g$ later.

The eigenoperator amplitudes in the amplitude channel corresponding to the lowest energy belong to the primary Higgs mode.
Even for $g \sim 0.3$, its energy lies about $10^{-7}W$ to $10^{-6}W$ below $2 \trueGap$.
The precise value depends on the discretization $N$.
Although we cannot provide a definite answer, a careful analysis of the effects of different discretizations suggests that the mode's binding energy might be exponentially small, cf.~\cref{app:finite_size}.

The results on the sc and bcc lattice are symmetric or antisymmetric for $\nu^{(1)}_j$ about $\varepsilon = \Ef = 0$.
On the fcc lattice, however, the amplitudes are skewed toward $\varepsilon < \Ef$ because 
of the absence of particle-hole symmetry.
Changing the Fermi level (not shown) produces the same effect for the other lattices as well.
Generally, the amplitudes on the bcc lattice are energetically narrower than the ones on the other lattices, 
which is certainly driven by the logarithmic singularity at the Fermi level.

The following discussion applies to all shown cases.
For small $g \lesssim 0.5$, the amplitudes $\alpha_j^{(1)}$ of the pair creation operators 
display a peak around the Fermi edge with a deep dip at  $\varepsilon=\mu$ where the amplitude vanishes.
Note that we chose the discretization such that it avoids $\varepsilon=0$.
Thus, we do not exactly evaluate the point $\varepsilon=\mu$,
but consider a point slightly above and another slightly below.
Extrapolating to finer and fine discretization shows that  $\alpha_j^{(1)} = 0$ at $\varepsilon=\mu$
is to be expected in the continuum limit.

Additionally, side peaks appear at $\mu \pm \debye$, likely due to the discontinuous cutoff of the interaction~\eqref{eqn:g}.
A possible explanation is found by computing $[\hamiltonian, \mathfrak{A}_\mathrm{Higgs}^{(n)}]$ and consequently applying Wick's theorem to reach bilinear operators. The result entails the expression
\begin{align}
\label{eqn:commutator_contribution}
    T_1 (\varepsilon) \coloneqq \sum_{\vq} \frac{g(\varepsilon_{\vq},\varepsilon)}{N} \frac{E_{\vq} \nu_{\vq}^{(n)}}{\Delta_{\vq}} = \int g(\varepsilon, \varepsilon') \rho(\varepsilon') \frac{E(\varepsilon') \nu_{\varepsilon'}^{(n)}}{\Delta (\varepsilon')} \md \varepsilon'.
\end{align}
The details are presented in \cref{app:commutators}.
If we used the BCS approximation $g=\mathrm{const}$, the above integrals would be independent of $\varepsilon$.
In our calculation, however, $g(\varepsilon, \varepsilon')$ restricts the integration domain 
to a small range of $\pm \debye$ around $\varepsilon$. 
This restriction implies that the Fermi level enters the integration domain only at $\varepsilon = -\debye$.
For small $g$, $T_1$ will be particularly large at $\varepsilon = \pm \debye$ because the integration domain includes only 
contributions of $\nu_{\varepsilon}$ of the same sign.
We expect that in the case of dispersive phonons, the boundaries would not be sharp anymore, so that these lobes are smeared out and would not appear.

As $g$ increases, the side peaks smear out, and the dip at the Fermi edge widens.
At $g \gtrsim 0.9$, the main peak has fully disappeared, and the amplitudes 
display two smooth peaks, one below and one above the Fermi level.
Increasing $g$ further shifts these peaks away from the Fermi level and widens them
but does not induce further qualitative changes.
The amplitudes $\nu_j^{(1)}$ of the number operators are antisymmetric about $\varepsilon = \Ef$ if the DOS is symmetric (sc and bcc lattices).
For small $g$, they are essentially $0$ except very close to the Fermi level.
There, they quickly rise to $1$ for $\varepsilon < \Ef$ and consequently rapidly fall off to $-1$.
This feature also widens as $g$ increases.

Next, we come back to the relation between the $\alpha$ and $\nu$ operator amplitudes.
We observe that
\begin{equation}
\label{eqn:alpha_nu_relation}
    \alpha_{\vk} = - \nu_{\vk} \frac{\dispersion{\vk} - \mu}{\Delta_{\vk}}
\end{equation}
holds up to $\mathcal{O}(10^{-9})$ for the fcc lattice and even up to $\mathcal{O}(10^{-13})$ for the particle-hole symmetric cases.
This remarkable precision cannot be accidental, but calls for an analytical justification.
Here, we briefly provide such an analytical argument.
On the level of bilinear operators, the operators
\begin{equation}
    \mathfrak{C}_{\vk} \coloneqq \Delta_{\vk} \left( \opSC{\vk} + \opSCd{\vk} \right) 
    + (\dispersion{\vk} - \mu) \left( \ladder[\dagger]{\vk}{\uparrow} \ladder{\vk}{\uparrow}
    + \ladder[\dagger]{-\vk}{\downarrow} \ladder{-\vk}{\downarrow}\right)
\end{equation}
are conserved quantities.
This is proven by explicitly computing $[\hamiltonian, \mathfrak{C}_{\vk}]$ and applying Wick's theorem afterwards as 
presented in \cref{app:commutators}.
The same set of operators can also be obtained by using linearized equations of motion~\cite{tsuji2015}.
Formally, these operators are eigenoperators of the Liouvillian with the eigenvalue $0$, i.e., 
$[ \hamiltonian, \mathfrak{C}_{\vk}] = 0$.

In the framework of the iEoM, we have to deal with $3N$-dimensional vectors that encode the operator amplitudes of every operator 
of the operator basis $\mathfrak{B}$. The conserved operator $\mathfrak{C}_{\vk}$ is represented by
\begin{equation}
    \vecstyle{c}_{\vk} = \vecstyle{e}_{\vk} \otimes \begin{pmatrix} \Delta_{\vk} \\ \dispersion{\vk} - \mu \\ 0 \end{pmatrix},
\end{equation}
where $\otimes$ means the direct product or Kronecker product and $\vecstyle{e}_{\vk}$ is a unit vector in the Brillouin zone,
i.e., for $N$ sites, this is a $\mathbb{R}^N$.
Consequently, since $[ \hamiltonian, \mathfrak{C}_{\vk}] = 0$, $\vecstyle{c}_{\vk}$ lies
in the kernel of the dynamical matrix~\eqref{eqn:dynmat}.
Moreover, it is easy to check that $\vecstyle{c}_{\vk}$ also lies in the kernel of the norm matrix $\mathcal{N}$.
Hence, the dynamics of the system completely decouple from any vector $\vecstyle{v} \in \mathrm{span}\left\{ \vecstyle{c}_{\vk} \, \middle\vert \, \vk \in 1^\mathrm{st}\, \mathrm{BZ} \right\}$.
Importantly, both $\dynamicalMatrix$ and $\normMatrix$ are Hermitian.
Hence, any excitation at finite energy must be orthogonal to their joint kernel, i.e., perpendicular to all $\vecstyle{c}_{\vk}$.
In the Higgs sector, this condition implies precisely the relation \eqref{eqn:alpha_nu_relation}. This concludes the
analytic argument.

Physically, the Higgs mode describes fluctuations of the magnitude of the order parameter $\Delta$.
As $\Delta$ directly enters the expectation values of the pair creation and of the number operators, 
it comes as no surprise that the eigenoperators of the Higgs modes comprise both pair creation/annihilation and number operators.
The relation between these two classes of operators is given by \eqref{eqn:alpha_nu_relation} as derived analytically above.

The amplitudes $\psi_j^{(1)}$ of the primary phase mode are shown in \cref{fig:primary_operators}(c).
Contrary to $\alpha_j^{(1)}$, these amplitudes are finite at $\varepsilon=\Ef$.
In fact, for $g \lesssim 0.9$, $\psi_j^{(1)}$ exhibits a pronounced peak around the Fermi level.
Increasing $g$ further causes the peak to split into two side peaks, leaving a finite minimum at the Fermi level itself.
Yet, the minimum values do not vanish.
In contrast to the Higgs mode, the phase mode does not affect the amplitude of $\Delta$, but its phase.
The phase of the order parameter does not affect the expectation values of the number operators at all.
Consequently, no number operators appear in the eigenoperators of the phase modes.
Conversely, the complex phase of $\expec{ \ladder{-k}{\down} \ladder{k}{\up} }$ does change.
Thus, it is natural to find that the $\opPhase{\epsilon}$-operators are the constituents of the 
phase eigenoperators. They measure the imaginary part of the expectation values $\expec{ \ladder{-k}{\down} \ladder{k}{\up} }$.

\begin{figure}
    \centering
    \includegraphics{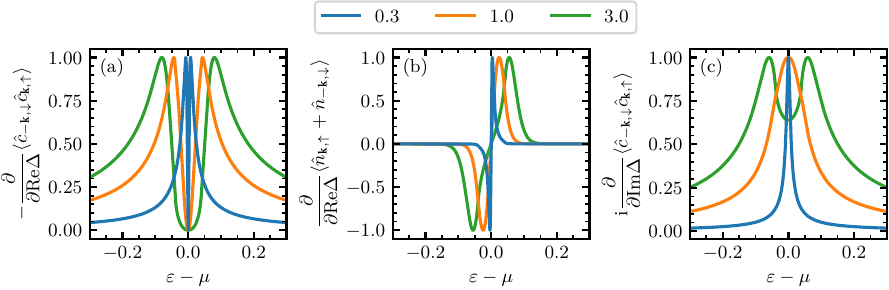}
    \caption{Plots of the derivatives of the expectation values, see \cref{eqn:deriv_expecs}, on the bcc lattice for $\Ef=0$.
    The colors represent different values of $g$.  The curves are normalized such that their maximum is $1$.}
    \label{fig:expectation_values}
\end{figure}

Above, we argued on a qualitative level with the relation between operators and expectation values.
This can be made more concrete, though not quantitative, by the following reasoning.
The concept of the collective excitations is that the Higgs modes change the amplitude of $\Delta$.
Given that we use the phase gauge in which $\Delta$ is real, this refers to the real part of $\Delta$.
In contrast, the phase modes change the imaginary part.
The following derivatives capture the leading effect of changes of $\Delta$ on the expectation values
\begin{subequations}
\label{eqn:deriv_expecs}
\begin{align}
    \frac{\partial}{\partial \Re \Delta} \langle \hat{c}_{-\mathbf{k},\downarrow} \hat{c}_{\mathbf{k},\uparrow} \rangle &=
        -\frac{(\varepsilon_{\vk} - \mu)^2}{2 E_{\vk}^3}, \label{eq:a}\\
    \frac{\partial}{\partial \Re \Delta} \langle \hat{n}_{\mathbf{k},\uparrow} + \hat{n}_{-\mathbf{k},\downarrow} \rangle &=
        \frac{(\varepsilon_{\vk} - \mu) \Delta_{\vk}}{E_{\vk}^3}, \label{eq:b}\\
\label{eqn:deriv_phase}
    \frac{\partial}{\partial \Im \Delta} \langle \hat{c}_{-\mathbf{k},\downarrow} \hat{c}_{\mathbf{k},\uparrow} \rangle \Big\vert_{\Im \Delta = 0} &=
        - \frac{\im}{2 E_{\vk}}.
\end{align}
\end{subequations}
These functions are plotted for the bcc lattice in \cref{fig:expectation_values}.
A comparison to the operator amplitudes displayed in \cref{fig:primary_operators} shows that they
resemble each other closely. Note that the results of \eqref{eq:a} and \eqref{eq:b} even fulfill the relation
\eqref{eqn:alpha_nu_relation} if we take the derivatives of the expectation values for operator amplitudes
and add the expectation value of the pair creation as well.

In the weak-coupling regime, the order parameter is confined to a narrow region around the Fermi edge.
Therefore, changes in the order parameter affect essentially only the immediate vicinity of the Fermi level.
Thus, narrow structures occur in this energy region.
In the strong-coupling regime, the expectation values change little near the Fermi level. 
Instead, the dominant effects result from contributions from additional momenta (or equivalently, energies).
Thus, we find the maxima in the amplitudes shifted away from the Fermi level.
Additionally, the understanding of the operator amplitudes via the expectation values conveniently explains 
$\alpha_j = 0$ at $\varepsilon = \mu$.
This phenomenon occurs because the derivative of the expectation value rigorously vanishes at this point.

\begin{figure}
    \centering
    \includegraphics{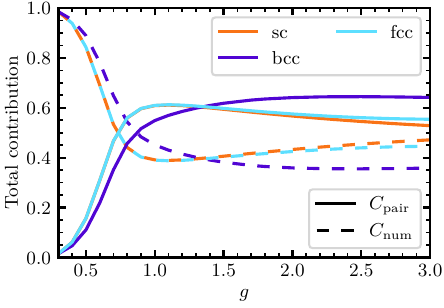}
    \caption{The contributions of the pair creation/annihilation operators (see \cref{eqn:cpair}, solid lines) 
    and of the number operators (see \cref{eqn:cnum}, dashed lines) to the primary Higgs mode as a function 
    of the interaction strength $g$.
    For the sc (orange lines) and fcc (cyan lines) lattices, the quantities are within line width the same for $g \lesssim 1.5$.
    The contributions for the bcc lattice (purple lines) are very similar, but quantitatively distinct.}
    \label{fig:primary_mode_integrals}
\end{figure}

Next, we return to the relative importance of pair creation/annihilation operators and 
number operators in the Higgs eigenoperators. The magnitudes of the amplitudes of the former are
denoted $|\alpha_j|$ and of the amplitudes of the latter $|\nu_j|$.
We define the the total contribution to $\vecstyle{\chi}^{(1)}$ of the pair creation/annihilation operators
\begin{subequations}
\begin{equation}
\label{eqn:cpair}
    C_\mathrm{pair} \coloneqq \frac{1}{|\vecstyle{\chi}^{(1)}|^2} \sum_j |\alpha_j^{(1)}|^2
\end{equation}
and the total contribution of the number operators
\begin{equation}
\label{eqn:cnum}
    C_\mathrm{num} \coloneqq \frac{1}{|\vecstyle{\chi}^{(1)}|^2} \sum_j |\nu_j^{(1)}|^2.
\end{equation}
\end{subequations}
Note that $C_\mathrm{pair} + C_\mathrm{num} = 1$ by construction.
In this way, $C_\alpha$ is a measure of which kinds of operators excite the primary Higgs mode.
These quantities are depicted in \cref{fig:primary_mode_integrals}.
Of course, this could be generalized to the secondary modes as well.

The solid curves represent $C_\mathrm{pair}$ while the dashed curves mark $C_\mathrm{num}$.
The colors distinguish the three lattices.
For $g\lesssim 1.5$, the contributions on the sc and fcc lattices are identical within line width.
Nonetheless, the quantities behave qualitatively the same even on the bcc lattice.
We ascribe the quantitative difference to the presence of the logarithmic divergence.

For small $g \lesssim 0.5$, the primary Higgs mode is primarily excited by the number operators, i.e., then, this mode consists mainly of electron-hole fluctuations.
Only for stronger interactions, the pair creation/annihilation operators become important and 
eventually even dominant.
Recent studies investigated higher-harmonic generation in resonance with the Higgs mode in superconductors.
That is, the incoming laser light satisfies $2\omega_\mathrm{Laser} = \omega_\mathrm{Higgs}$ and generates resonant output at $3\omega_\mathrm{Laser}$ or even $5\omega_\mathrm{Laser}$.
The studies showed that these higher-harmonic generations are mainly driven by charge-density fluctuations, and that the pair-breaking contribution is minute~\cite{cea2016,schwarz2020b,reinhoffer2022}.
This observation aligns well with the present finding. 

\begin{figure}[!t]
    \centering
    \includegraphics{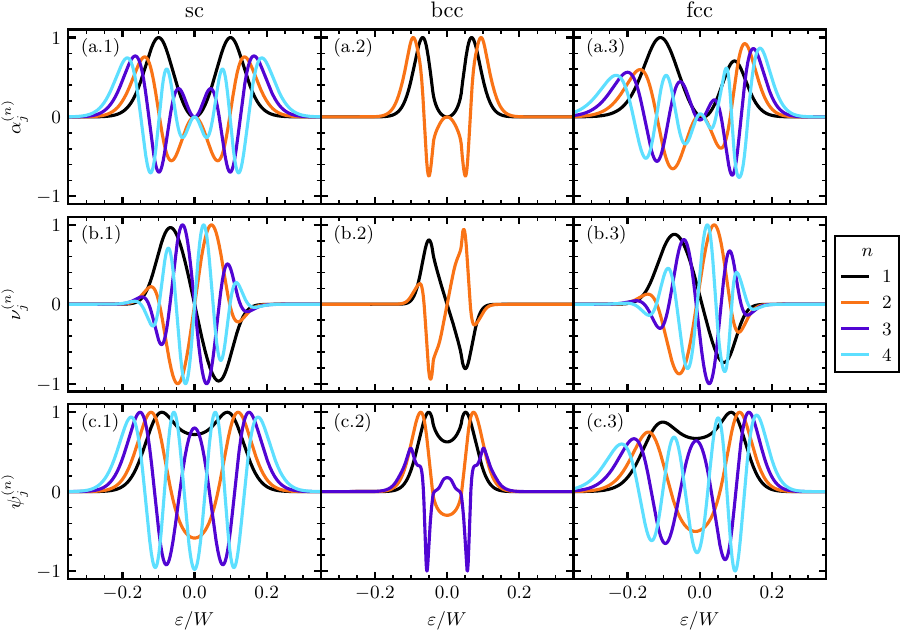}
    \caption{The operator amplitudes (see \cref{eqn:excite_higgs,eqn:excite_phase}) for the Higgs and phase modes on the (1) sc, (2) bcc, and (3) fcc lattice for $\Ef = 0$. The colors represent the $n$-th respective mode, see legend.}
    \label{fig:secondary_operators}
\end{figure}

Last but not least, we turn toward the eigenoperators of the secondary modes.
Their amplitudes for $g=3$ are plotted in \cref{fig:secondary_operators}.
The results for different values of $g$ are qualitatively identical; we illustrate the results for a large value of $g$ for clarity.
The columns of panels show the data for the three lattices. 
The rows depict the operator amplitudes of the (a) $\opAmplitude{\varepsilon}$, (b) $\opNumber{\varepsilon}$, and (c) $\opPhase{\varepsilon}$ operators. 
The amplitudes of the primary mode are shown in black.
The colored curves correspond to all secondary modes found in the respective systems for $g=3$.

For the particle-hole symmetric systems, $\alpha_j^{(n)}$ and $\psi_j^{(n)}$ are symmetric and $\nu_j^{(n)}$ is antisymmetric about $\varepsilon = \Ef$. 
On the fcc lattice, the amplitudes are skewed because the DOS is not symmetric, but they still vanish at $\varepsilon=\mu$.
The amplitudes $\alpha_j^{(n)}$ and $\nu_j^{(n)}$ vanish around the Fermi level, while $\psi_j^{(n)}$ remains finite.
The amplitudes on the bcc lattice exhibit a few kinks that are likely related to the logarithmic divergence of the DOS.

The operator amplitudes of the individual modes exhibit a clear and distinct behavior.
Not counting the zero at $\varepsilon = \mu$, the primary modes have no roots.
The second mode has 2 roots, the third mode has 4, and the fourth mode has 6.
Thus, we conjecture that the number of roots is given by $2 (n-1)$ for the $n$-th mode.
Such behavior is strongly reminiscent of the wave functions of bound states in basic one-dimensional 
quantum mechanical systems with attractive potentials
such as the harmonic oscillator or the radial problem for the eigenstates of the Hydrogen atom.
It is tempting to claim a Sturm–Liouville oscillation theorem for operator amplitudes.

Furthermore, we recall that the amplitudes depend on the momenta only via the electronic energies $\varepsilon$
so that they have the same point-group symmetries as the lattice. Thus, this applies also to the 
eigenoperators and their operator amplitudes.
Thus, the secondary modes are (extended) $s$-wave modes.
Ref.~\cite{althuser2025} already hypothesized that the secondary excitations are related to a 
sort of principal quantum number. The analysis here corroborates this view.

\section{Conclusion and outlook}
\label{sec:conclusion}

We investigated superconductivity on three-dimensional Bravais lattices as it is induced by
electron-phonon interaction. The latter is mapped to an all-electron model by a systematically
controlled change of basis, a continuous unitary transformation. This reveals that the attractive
interaction is active as long as the energy difference of the scattered electron is 
smaller in magnitude than the phonon energy. 

This systematically derived effective model is investigated for collective excitations in the Higgs and in the phase channel.
The leading, primary Higgs and phase modes could be identified readily. 
For moderate to strong coupling, further secondary modes were found. We determined the eigenoperators, i.e., the operators that excite single collective modes specifically. 
The operator amplitudes of these modes were computed.

The main results of our study are threefold.
First, on the mean-field level, we saw that the minimum of the quasiparticle dispersion does not need to lie precisely at the Fermi edge $\varepsilon = \mu$.
Instead, the minimum can shift away from the Fermi edge if $|\varepsilon - \mu|$ increases more slowly than $\Delta(\varepsilon)$ decreases.
This phenomenon has been found in a continuum model as well~\cite{althuser2025}.

Second, we investigated how the spectra of collective excitations presented in Ref.~\cite{althuser2025} are modified on lattices. A particular focus was on the existence of the secondary amplitude and phase modes.
We can report essentially the same situations as in the reference: the primary Higgs mode is located at the lower edge of the quasiparticle continuum $2\trueGap$ 
only in the weak-coupling limit.
At intermediate coupling strengths, the mode detaches itself and becomes dissipationless because it can no longer decay into particle-hole pairs.
This behavior stems from the nontrivial momentum dependence of the pairing potential as observed in other studies~\cite {barankov2007,althuser2025,althuser2025a} as well.
Increasing the coupling further causes secondary amplitude and phase modes
to emerge from the quasiparticle continuum.
These unconventional additional excitations appear at regular intervals of the energy gap.
Moreover, these intervals in units of the Debye frequency $\debye$ are not sensitive 
to the specific details of the model: we observed an additional mode approximately every $0.63 \debye$ upon increasing the system's energy gap
by increasing the coupling.
A nearly identical number is obtained for a parabolic dispersion including the Coulomb interaction~\cite{althuser2025}. Thus, this appears to be a universal property.

Lastly, we extended the understanding of the secondary excitations.
In the framework of iterated equations of motion, we analyzed the Fourier-transformed Green's functions 
and identified the eigenoperators of the collective modes. These are the operators that excite or de-excite only one specific mode. The prefactors defining the eigenoperators, 
dubbed operator amplitudes, have been computed explicitly.
For the primary Higgs and phase modes, these amplitudes strongly resemble the derivatives of the corresponding expectation values, cf.~\cref{eqn:deriv_expecs}.
We derived an analytical relation between the charge and the pair creation or annihilation
amplitudes of the Higgs modes on the level of the employed approximations.

For the secondary modes, the operator amplitudes display nodal structures
strongly reminiscent of the oscillation theorem for one-dimensional quantum systems.
The amplitudes of the secondary Higgs and phase modes have two roots, 
the third ones have four, the fourth ones have six, and so on.
The secondary modes are similar to Bardasis-Schrieffer modes~\cite{bardasis1961,vaks1962,vaks1962a,barlas2013,bohm2014,sun2020a,muller2021,hackner2023}, which correspond to different angular momentum $L$. However, we emphasize that the modes 
identified here have the same symmetries as the underlying lattice.
In particular, no finite momentum or angular momentum is involved. 
They rather correspond to different ``radial'' modes in energy. This distinguishes them clearly from Bardasis-Schrieffer modes.

Clearly, our findings call for many future follow-up investigations.
On the theoretical side, one can envision extending the present calculations to
collective excitations with finite angular momentum. To this end, the set of
basis operators has to be changed such that they add or subtract a finite 
angular momentum to the system. Similarly, one can design a modified operator
basis comprising operators at finite total momentum. This will allow one to 
address the dispersion of the collective modes of Higgs and phase character.

Another intriguing route of theoretical studies consists of developing 
formulae or simulations for potential experimental probes. A fascinating
objective concerns the probe of the eigenoperators, i.e., of the structure
of the collective modes in momentum space or in real space.

On the experimental side, the most important route is the
realization and verification of the collective modes predicted in our simulations.
Promising candidate material should display a large superconducting gap
relative to the energy of the exchanged boson. Such a situation is strongly favored
by a large electronic density of states at the Fermi level.

The side minima and the W-shape of the quasiparticle dispersion are verifiable 
by measuring the quasiparticle density-of-states, for instance by means of tunneling spectroscopy~\cite{hanaguri2010,wen2025}, or by directly accessing the order parameter in angle-resolved photoemission spectroscopy~\cite{okazaki2012,sobota2021}.
A W-shape entails additional flat regions in the dispersion, which in turn imply additional peaks in the corresponding DOS.

The spectra of collective excitations presented here do not hinge on 
any finite momentum or angular momentum. Therefore, they should be accessible 
by means of any conventional method capable of identifying the primary Higgs mode.
The established examples are Raman spectroscopy~\cite{sooryakumar1980,littlewood1981,littlewood1982,devereaux2007,cea2014,glier2025} or Terahertz spectroscopy~\cite{matsunaga2014,katsumi2018,chu2020,katsumi2020,reinhoffer2022}.
Recently, even a momentum-resolved study of the condensate dynamics was suggested~\cite{schwarz2020}.

Summarizing, the finding of secondary modes and the computation of their internal structures in the operator amplitudes sets the stage for a very rich field of studies on superconductivity. Competing ordering phenomena can be considered on equal footing, giving even access to the phase transitions~\cite{althuser2024}.

\ack{We would like to thank D.-B. Hering, I. M. Eremin, and J. Stolze for very helpful discussions.}

\funding{This research was partially funded by the MERCUR Kooperation in project KO-2021-0027.}

\roles{
Joshua Alth\"user carried out the calculations and implemented the numerical
simulations, created the figures, and wrote large parts of the manuscript. G\"otz S.~Uhrig
initially had suggested the approach and supervised the project. 
Both authors interpreted the results for their
physical implications and edited the manuscript.}

\data{The data will be made available prior to publishing at TUDOData.}

\appendix

\section{Computing specific mode-exciting eigenoperators}
\label{app:computing_operators}

Following Ref.~\cite{althuser2024}, the Fourier-transformed Green's functions are expressed in terms of squared complex frequencies $z^2$ and blocks of the matrices $\dynamicalMatrix$ and $\normMatrix$ as long as the appearing expectation values are real.
This condition is fulfilled in the entirety of this article.

In particular, we split the operator of the basis into two sets, the Hermitian operators 
and the anti-Hermitian ones. The former are denoted by $\hat{X}_j$ and the latter
by $\hat{P}_j$. Then, we define
\begin{subequations}
\begin{align}
    \left( \mathcal{K}_+ \right)_{ij} &\coloneqq \opprod{\hat{X}_i}{[\hamiltonian, \hat{X}_j]}, \\
    \left( \mathcal{K}_- \right)_{ij} &\coloneqq \opprod{\hat{P}_i}{[\hamiltonian, \hat{P}_j]}, \\
    \left( \mathcal{L} \right)_{ij}   &\coloneqq \opprod{\hat{X}_i}{\hat{P}_j}.
\end{align}
\end{subequations}
For this article, there are $2N$ Hermitian $\hat{X}$-operators, namely $\hat{X}_i = \opAmplitude{\varepsilon_i}$ and $\hat{X}_{i+N} = \opNumber{\varepsilon_i}$ for $i < N$.
Analogously, $\hat{P}_i$ represents the anti-Hermitian operators $\opPhase{\varepsilon_i}$.

Then, the amplitude of Higgs Green's function is given by
\begin{equation}
    \greenMat{Higgs} = \vecstyle{x}^\dagger \mathcal{T}_X \frac{1}{z^2 - \check{M}_P} \mathcal{T}_X^\dagger \vecstyle{x}
\end{equation}
with $z\coloneqq \omega + \im 0^+$ and the definitions
\begin{subequations}
\begin{align}
    \mathcal{T}_X &\coloneqq \left( \mathcal{L}^\dagger \mathcal{K}_+^{-1} \mathcal{L} \right)^{-1/2} \mathcal{L}^\dagger \\
    \check{M}_P   &\coloneqq \left( \mathcal{L}^\dagger \mathcal{K}_+^{-1} \mathcal{L} \right)^{-1/2} \mathcal{K}_- \left( \mathcal{L}^\dagger \mathcal{K}_+^{-1} \mathcal{L} \right)^{-1/2}.
\end{align}
\end{subequations}
The vector $\vecstyle{x}$ has to be chosen such that $\mathfrak{A}_\text{Higgs}~\equiv \sum_i x_i \hat{X}_i$.
An analogous expression can be derived for $\greenMat{Phase}$ as well~\cite{althuser2024}.

The energy of subgap excitations is given by $\omega_0^2 - \check{M}_P = 0$, i.e., by the square root of the eigenvalues of $\check{M}_P$.
Computing the corresponding eigenvector allows us to identify which operators specifically excite this mode.
To this end, we diagonalize $\check{M}_P = \mathcal{U} \mathcal{D} \mathcal{U}^\dagger$
where the columns of $\mathcal{U}$ contain the eigenvectors $\vecstyle{u}_n$ and $\mathcal{D} = \text{diag}(\omega_n^2)$. Then
\begin{align}
    \greenMat{Higgs} &= \vecstyle{x}^\dagger \mathcal{T}_X^\dagger \mathcal{U} \frac{1}{z^2 - \mathcal{D}} \mathcal{U}^\dagger \mathcal{T}_X \vecstyle{x} 
    = \sum_{n} \frac{| \vecstyle{u}_n^\dagger \mathcal{T}_X \vecstyle{x} |^2}{z^2 - \omega_n^2}.
\end{align}
The vector $\vecstyle{\chi}^{(n)}$ is defined by
\begin{align}
    \label{eqn:def_chi}
    \mathcal{T}_X \vecstyle{\chi}^{(n)} = C_n \vecstyle{u}_n
\end{align}
and contains the sought information. The constant $C_n$ is arbitrary because it
merely becomes a factor in the final Green's function.
The entries of $\vecstyle{\chi}^{(n)}$ are the operator amplitudes discussed in the main text.
They tell us which linear superposition of $\hat{X}$-operators excites or de-excites 
the specific mode at energy $\omega_n$.
Concretely, the operator is given by $\mathfrak{A}_\mathrm{Higgs}^{(n)} = \sum_j {\chi}_j^{(n)} \hat{X}_j$ implying 
\begin{align}
    \greenMatSymbol_\mathrm{Higgs}^{(n)} (\omega) = \sum_j \frac{| \vecstyle{u}_j^\dagger \mathcal{T}_X \vecstyle{\chi}^{(n)} |^2}{z^2 - \omega_j^2} = \sum_j \frac{| \vecstyle{u}_j^\dagger C_n \vecstyle{u}_n |^2}{z^2 - \omega_j^2} = \frac{|C_n|^2}{z^2 - \omega_n^2}.
\end{align}
Thus, the operator $\mathfrak{A}_\mathrm{Higgs}^{(n)}$ excites or de-excites only the mode with energy $\omega_n$. The normalization of $\vecstyle{\chi}^{(n)}$ can be modified
by adjusting $C_n$. For plotting purposes, we normalize the amplitudes such that 
$\max_j | \chi_j^{(i)}| = 1$.

Moreover, due to the choice of basis operators~\eqref{eqn:basis_operators}, we can associate the $j$-th entry of $\vecstyle{\chi}^{(n)}$ to a unique operator from \cref{eqn:basis_operators}.
For $j \leq N$, this operator is denoted $\hat{R}{(\varepsilon_j)}$.
For $j > N$, this operator is denoted $\hat{N}{(\varepsilon_{j-N})}$.
In this way, we determine the operator amplitudes as functions of energy.

Mutatis mutandis, one obtains the equivalent for $\hat{P}$-type operators via
\begin{align}
    \label{eqn:def_psi}
    \mathcal{T}_P \vecstyle{\psi}^{(n)} = \tilde{C}_n \vecstyle{v}_n.
\end{align}
Here, $\vecstyle{v}_n$ is the corresponding eigenvector of $\check{M}_X$ and
\begin{subequations}
\begin{align}
    \mathcal{T}_P &\coloneqq \left( \mathcal{L} \mathcal{K}_-^{-1} \mathcal{L}^\dagger \right)^{-1/2} \mathcal{L} \\
    \check{M}_X   &\coloneqq \left( \mathcal{L} \mathcal{K}_-^{-1} \mathcal{L}^\dagger \right)^{-1/2} \mathcal{K}_+ \left( \mathcal{L} \mathcal{K}_-^{-1} \mathcal{L}^\dagger \right)^{-1/2}.
\end{align}
\end{subequations}
Here, the $j$-th entry of $\vecstyle{\psi}^{(n)}$ corresponds to $\hat{I}{(\varepsilon_j)}$.

In general, the matrix $\mathcal{L}$ is not square and will have a large kernel.
Therefore, it is mathematically not guaranteed that a solution to \cref{eqn:def_chi,eqn:def_psi} exists.
However, if such a solution does not exist, i.e., $\vecstyle{u}_n \in \text{ker}(\mathcal{T}_X)$ or $\vecstyle{v}_n \in \text{ker}(\mathcal{T}_P)$, the eigenvalue would not contribute to the spectral functions and would not be identified by our method anyway.
Thus, the inverse is needed only in the subspace where it exists, so that
the solvability of \cref{eqn:def_chi,eqn:def_psi} is not an issue.
In practice, we find $|| \check{M}_P \mathcal{T}_X \vecstyle{\chi}^{(n)} - \omega_n^2 \vecstyle{u}_n || \lesssim 10^{-14}$ and similar for $\vecstyle{\psi}^{(n)}$.
Thus, the numerics find the appropriate solution without problems.

In general, eigenvectors corresponding to well-separated eigenvalues can be efficiently computed using standard Lanczos methods. But this condition is obviously not satisfied for modes close to the continuum.
In this case, it is best to diagonalize the $\check{M}$-matrices completely.
While this is computationally much more demanding, it can still be carried out on current workstations in a few hours.

\section{Commutators with the Hamiltonian}
\label{app:commutators}

In this appendix, we compute the relevant commutators with the Hamiltonian~\eqref{eqn:h}.
We define the abbreviations $\xi_{\vk} \coloneqq \dispersion{\vk} - \mu$ and $\opf{\vk} \coloneqq \opSC{\vk}$, and evaluate
\begin{subequations}
\begin{align}
    [\hamiltonian, \opf{\vk} + \opf[\dagger]{\vk}] = &-2 \Big[
        \xi_{\vk} \opf{\vk} + \sum_{\vq} \frac{g(\vk, \vq)}{N} \Big( 1 - \opn{\vk}{\up} - \opn{-\vk}{\down} \Big) \opf{\vq}
    \Big] - \hc  \\
    [\hamiltonian, \opn{\vk}{\up} + \opn{-\vk}{\down}] = &- 4 \sum_{\vq} \frac{g(\vk, \vq)}{N} \left( 
        \opf[\dagger]{\vk} \opf{\vq} - \hc
    \right).
\end{align}
\end{subequations}
Next, we apply Wick's theorem to these expressions with respect to the
mean-field ground state and omit the quartic terms. This is justified 
because the resulting operators are used to determine the norm and the dynamic matrix
where the expectation values with respect to the ground state of the bilinear
mean-field Hamiltonian enter. Thus, the application of Wick's theorem is warranted.
Inserting the mean-field expectation values
\begin{subequations}
\begin{align}
    \expec{\opf{\vk}} &= - \frac{\Delta_{\vk}}{2 E_{\vk}} \\
    \expec{1 - \opn{\vk}{\up} - \opn{-\vk}{\down}} &= \frac{\xi_{\vk}}{E_{\vk}}
\end{align}
\end{subequations}
eventually yields
\begin{subequations}
\begin{align}
\label{eqn:comm_H_f}
    [\hamiltonian, \opf{\vk} + \opf[\dagger]{\vk}] &= -2 \Big[
        \xi_{\vk} \opf{\vk} + \frac{\xi_{\vk}}{E_{\vk}} \sum_{\vq} \frac{g(\vk, \vq)}{N} \opf{\vq}
    \Big] - \hc \\
\label{eqn:comm_H_n}
    [\hamiltonian, \opn{\vk}{\up} + \opn{-\vk}{\down}] &= 2 \Big[
        \Delta_{\vk} \opf{\vk} + \frac{\Delta_{\vk}}{E_{\vk}} \sum_{\vq} \frac{g(\vk, \vq)}{N} \opf{\vq}
    \Big] - \hc \\
    \Rightarrow [\hamiltonian, \opn{\vk}{\up} + \opn{-\vk}{\down}] &= -\frac{\Delta_{\vk}}{\xi_{\vk}} [\hamiltonian, \opf{\vk} + \hc] .
\end{align}
\end{subequations}
This implies that the operators $\mathfrak{C}_{\vk} \coloneqq \xi_{\vk} (\opn{\vk}{\up} + \opn{-\vk}{\down}) + \Delta_{\vk} (\opf{\vk} + \opf[\dagger]{\vk})$ are constants of motion for each value of $\vk$.
It should be kept in mind that the derivation above omits normal-ordered quartic terms.
The same relation can be found using linearized equations of motion as done in Ref.~\cite{tsuji2015}.

By using \cref{eqn:comm_H_f,eqn:comm_H_n}, we obtain
\begin{subequations}
\begin{align}
    [\hamiltonian, \mathfrak{A}_\mathrm{Higgs}^{(n)}] 
    &\stackrel{\phantom{\text{\cref{eqn:alpha_nu_relation}}}}{=} 2 \sum_{\vk} \left[ \Delta_{\vk} \nu_{\vk}^{(n)} - \xi_{\vk} \alpha_{\vk}^{(n)} 
    + \sum_{\vq} \frac{g(\vq,\vk)}{N} \left( \frac{\Delta_{\vq} \nu_{\vq}^{(n)}}{E_{\vq}}
    - \frac{\xi_{\vq} \alpha_{\vq}^{(n)}}{E_{\vq}} \right) \right] \opf{\vk} - \hc \\
    \label{eqn:comm_H_higgs}
    &\stackrel{\text{\cref{eqn:alpha_nu_relation}}}{=} 2 \sum_{\vk} \bigg[ \underbrace{\frac{E_{\vk}^2 \nu_{\vk}^{(n)}}{\Delta_{\vk}}}_{\coloneqq T_0}
    + \underbrace{\sum_{\vq} \frac{g(\vq,\vk)}{N} \frac{E_{\vq} \nu_{\vq}^{(n)}}{\Delta_{\vq}}}_{\coloneqq T_1} \bigg] \opf{\vk} - \hc.
\end{align}
\end{subequations}
The first term $T_0$ represents only a rescaled version of the input $\nu_{\vk}^{(n)}$.
The second term $T_1$ is more interesting.
Let us first employ the BCS approximation that $g(\vk,\vq)$ is a constant in some region around the Fermi level, and we assume that the system displays particle-hole symmetry.
Then, we have that $\Delta_{\vq} = \mathrm{const}$ and that $E_{\vq}$ is invariant under
sign flip $\varepsilon \to -\varepsilon$.
Furthermore, we saw above that $\nu_{\vq}$ is antisymmetric, i.e., it acquires a sign flip 
for $\varepsilon \to -\varepsilon$. Combining these facts implies that $T_2 = 0$.
But the same line of reasoning does not hold for the interaction~\eqref{eqn:intro_g}.
Thus, the term $T_2$ is a direct consequence of the more systematic interaction
\eqref{eqn:intro_g}.

\section{The third amplitude mode on the fcc lattice}
\label{app:third_fcc_mode}

\begin{figure}
    \centering
    \includegraphics{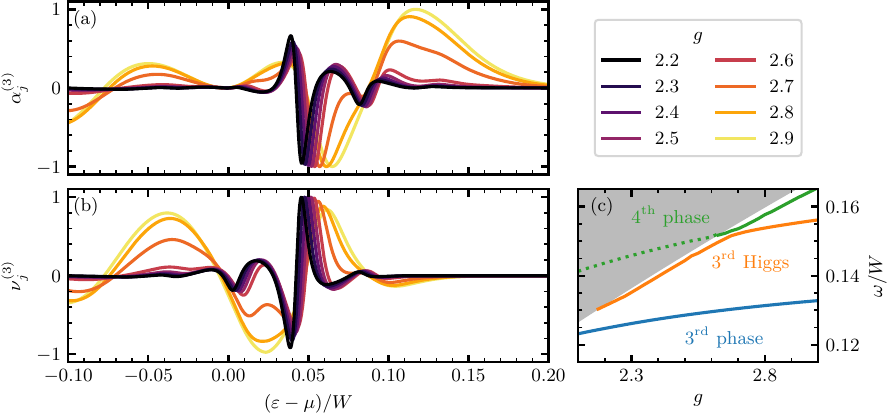}
    \caption{(a,b) Operator amplitudes (see \cref{eqn:excite_higgs}) of the eigenoperator 
    of the third Higgs mode on the fcc lattice at $\Ef = -0.5W$ for various $g$.
    The operator amplitudes evolve as $g$ is varied.
    (c) Energy of the collective excitations found in \cref{fig:heatmap_g_EF}.
    The plot is zoomed in on the region where the third Higgs mode emerges.
    The shading marks the two-particle continuum. The third phase mode is depicted as well.
    }
    \label{fig:fcc_third_mode}
\end{figure}

As mentioned in the main text, the third amplitude mode on the fcc lattice with $\Ef = -0.5W$ behaves in a slightly peculiar way. The behavior is shown more clearly in \cref{fig:fcc_third_mode}(c).
The mode emerges at $g\approx 2.2$ from the continuum, but stays very close to it.
Only when the fourth phase mode emerges do the two modes experience level repulsion, and they undergo an anticrossing as can be discerned in the figure.
Only beyond the anticrossing, the third Higgs mode acquires significant spectral weight.

Figures \ref{fig:fcc_third_mode}(a,b) depict the operator amplitudes corresponding to this mode.
As $g$ increases, the modes evolve into one another, suggesting that this mode is truly the same for quite different values of $g$.
However, as long as the mode energy remains close to the continuum, 
the amplitudes are restricted to a narrow energy window around the Fermi level. 
This behavior points to a considerable hybridization of the third Higgs mode 
with nearby quasiparticle excitations.

\section{Analysis of the discretization}
\label{app:finite_size}

\begin{figure}
    \centering
    \includegraphics{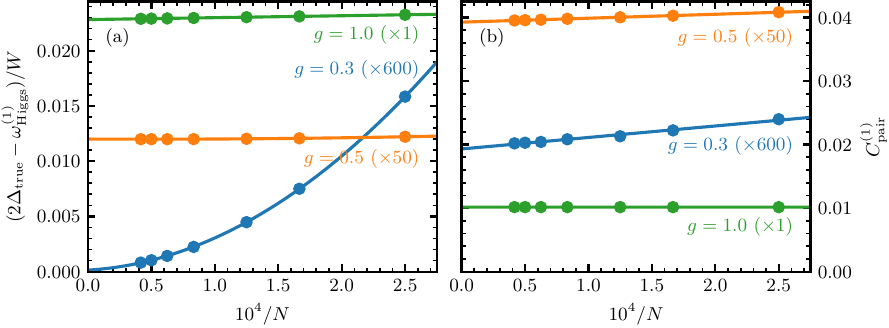}
    \caption{(a) Energetic spacing between the primary Higgs mode and the quasiparticle continuum as a function of the inverse number of discretization points $1/N$.
    (b) $C_\mathrm{pair}^{(1)}$, see \cref{eqn:cpair}, as a function of $1/N$.
    The shown data has been computed on the sc lattice at $\Ef=0$.
    The markers are the evaluated data points. 
    The data points were scaled by the factors given in the panels for clarity.
    The colors correspond to $g=0.3$ (blue, scaled by $600$), $g=0.5$ (orange, scaled by $50$), 
    and $g=1$ (green, not scaled). 
    The curves are least-squares fits to (a) a linear function and (b) a parabolic function.}
    \label{fig:finite_size}
\end{figure}

In this Appendix, we provide a brief overview of the effects of changing the number $N$ of discretization points.
The behavior is qualitatively the same in all considered cases, 
hence we restrict ourselves to the sc lattice with $\Ef=0$.
\Cref{fig:finite_size} shows (a) the energetic spacing between the primary Higgs mode and the quasiparticle continuum and (b) $C_\mathrm{pair}^{(1)}$, see \cref{eqn:cpair}, as a function of $1/N$.
The markers represent the evaluated data points, scaled by the indicated factors for clarity.
The colors correspond to $g=0.3$ (blue, scaled by $600$), $g=0.5$ (orange, scaled by $50$), and $g=1$ (green, not scaled).
The curves are least-squares fits to
\begin{subequations}
\begin{equation}
    (2 \Delta_\mathrm{true} - \omega_\mathrm{Higgs}^{(1)}) / W = a x^2 + bx + c
\end{equation}
and to
\begin{equation}
    C_\mathrm{pair}^{(1)} = d x + e,
\end{equation}
\end{subequations}
where $x \coloneqq 10^{4} / N$.
The fit parameters are given in \cref{tab:finite_size_fit}.

The left-most data points in \Cref{fig:finite_size} are at $N=20000$; the second ones are at $N=16000$; this is the discretization used in all the results shown in the main text.
Importantly, there is never a large discrepancy between the value for $N \to \infty$ obtained via the fit and the data points for $N=16000$.
The extrapolation for small $g$ indicates that the Higgs mode is detached from the continuum even for $g=0.3$, but by a very small binding energy, which might be exponentially small. 
However, the numerics cannot exclude the degeneracy of the Higgs mode and the lower continuum edge. 

\begin{table}[!h]
    \centering
    \begin{tabular}{c|ccc}
        $g$ & 0.3 & 0.5 & 1.0 \\
        \hline 
        $a$ & $(3.75 \pm 0.02) \times 10^{-6}$ & $\phantom{-}(1.068    \pm 0.002)   \times 10^{-6}$ & $\phantom{-}(5.382     \pm 0.010)     \times 10^{-7}$ \\
        $b$ & $(1.10 \pm 0.05) \times 10^{-6}$ & $(-1.009  \pm 0.005)   \times 10^{-6}$ & $\phantom{-}(1.72706   \pm 0.00003)   \times 10^{-4}$ \\
        $c$ & $(2.4  \pm 0.3)  \times 10^{-7}$ & $\phantom{-}(2.40043 \pm 0.00003) \times 10^{-4}$ & $\phantom{-}(2.2813983 \pm 0.0000002) \times 10^{-2}$ \\
        $d$ & $(3.0  \pm 0.1)  \times 10^{-6}$ & $\phantom{-}(1.24    \pm 0.01)    \times 10^{-5}$ & $(-4.08     \pm 0.06)      \times 10^{-6}$ \\
        $e$ & $(3.22 \pm 0.02) \times 10^{-5}$ & $\phantom{-}(7.854   \pm 0.002)   \times 10^{-4}$ & $\phantom{-}(1.016337  \pm 0.000010)  \times 10^{-2}$ 
    \end{tabular}
    \caption{Best fit parameters for the fits in \cref{fig:finite_size}. The given numbers are not rescaled.}
    \label{tab:finite_size_fit}
\end{table}


\section{Densities of states}
\label{app:dos}
All DOS are given for a tight-binding model with isotropic nearest-neighbor hopping.
The abcissae in $\varepsilon$-space are chosen such that the logarithmic singularities are avoided.

The DOS of the sc lattice is given by~\cite{hanisch1997}
\begin{equation}
    \rho_\mathrm{sc}(\varepsilon) = \frac{1}{\pi} \int_{u_1}^{u_2} \frac{\md u}{\sqrt{1-u^2}} \rho_{\square}( \varepsilon +  u W / 2)
\end{equation}
with
\begin{equation}
    u_1 = \max (-1, -2 - 2 \varepsilon / W), \qquad 
    u_2 = \min (1, 2 - 2 \varepsilon / W).
\end{equation}
The DOS of the square lattice reads
\begin{equation}
    \rho_{\square} (\varepsilon) = \frac{2}{\pi^2 W} K \left( 1 - \frac{\varepsilon^2}{W^2} \right),
\end{equation}
where $K(x)$ denotes the complete elliptic integral of the first kind.

The DOS of the bcc lattice is given by~\cite{joyce1971,joyce1994}
\begin{equation}
    \rho_\mathrm{bcc} (\varepsilon) = - \frac{4}{ \pi^3 W} \Im \left[ z \left( K \left( \frac{1}{2} - \frac{1}{2} \sqrt{1 - z^2} \right) \right)^2 \right],
\end{equation}
where $z \coloneqq W / [\epsilon + \im 0^+]$.

The DOS of the fcc lattice is given by~\cite{joyce2011}
\begin{equation}
    \rho_\mathrm{fcc} (\varepsilon) = - \frac{8}{\pi^3 W} \Im \left[ \tilde{z} \left( 1+ \tilde{z} \right)^{-3/2} \left( 2 - \sqrt{ 1 - 3 \tilde{z}} \right) K^2 [k_- ] \right],
\end{equation}
where
\begin{equation}
    k_-^2 \coloneqq \frac{1}{2} - 2\tilde{z} \left( 1 + \tilde{z} \right)^{-3/2} - \frac{1}{2} \sqrt{1 - 3\tilde{z}} \left( 1 - \tilde{z} \right) \left( 1 + \tilde{z} \right)^{-3/2} 
\end{equation}
and $\tilde{z} = W / [(2 \varepsilon + W) + \im 0^+]$.

\printbibliography

\end{document}